\documentclass[a4paper,11pt]{article}
\usepackage{graphicx}
\usepackage{amsmath}
\usepackage{amssymb}
\usepackage{latexsym}
\usepackage[colorlinks]{hyperref}
\usepackage{color}
\usepackage{float}

\usepackage[textfont={footnotesize,sf},labelfont={color=blue,bf,sf},labelsep=endash]{caption}
\usepackage[position=top,labelfont={color=blue,bf,sf}]{subfig}
\usepackage{bm}

\newcommand{\bra}{\begin{array}}
\newcommand{\era}{\end{array}}
\newcommand{\beq}{\begin{equation}}
\newcommand{\eeq}{\end{equation}}
\newcommand{\beqar}{\begin{eqnarray}}
\newcommand{\eeqar}{\end{eqnarray}}

\def\BC{\bb C}
\def\_\BC{\bbi C}

\def\Tr {{\rm Tr}}

\def\( {\left(}
   \def\) {\right)}
\def\[ {\left[}
\def\] {\right]}
\def\no2 {{\textstyle{n\over 2}}}

\def\Tr {{\rm Tr}}

\newcommand{\om}{\omega}

\newcommand{\te}{\theta}

\newcommand{\al}{\alpha}

\newcommand{\da}{\dagger}

\newcommand{\ov}{\over}

\newcommand{\sq}{\sqrt}

\newcommand{\lb}{\label}

\begin{document}
\begin{titlepage}
\setcounter{page}{1}
\renewcommand{\thefootnote}{\fnsymbol{footnote}}

\begin{flushright}
%ucd-tpg: 07/12/2012\\
%arXiv:yymm.xxxx
\end{flushright}

\vspace{5mm}

\begin{center}
 {\Large \bf %Spin Hall Effect for
 Thermodynamics Properties of Confined Particles on Noncommutative Plane }\\

\vspace{5mm}

{\bf Rachid Hou\c{c}a}$^{a}$ and {\bf Ahmed Jellal\footnote{\sf
a.jellal@ucd.ac.ma}}$^{a,c}$

\vspace{5mm}

{$^{a}$\em Department of Physics,
Faculty of Sciences, Ibn Zohr University,}\\
{\em PO Box 8106, Agadir, Morocco}

{$^b$\em Saudi Center for Theoretical Physics, Dhahran, Saudi
Arabia}

{$^{c}$\em Theoretical Physics Group,  %Department of Physics,
Faculty of Sciences, Choua\"ib Doukkali University},\\
{\em PO Box 20, 24000 El Jadida, Morocco}

\vspace{3cm}

\begin{abstract}

We consider a system of $N$ particles living on the noncommutative plane
in the presence of a confining potential and study its thermodynamics properties.
Indeed, after calculating the partition function, we determine the corresponding
internal energy and heat capacity where different corrections are obtained.
In analogy with the magnetic field case, we define an effective magnetization
and study its susceptibility in terms of
the noncommutative parameter $\theta$. By introducing the chemical potential, we investigate the Bose-Einstein condensation
for the present system.
Different limiting cases related to the temperature and $\theta$ will be analyzed as well
as some numerical illustration will be presented.

\end{abstract}
\end{center}
\vspace{3cm}


\end{titlepage}

%%%%%%%%%%%%%%%%%%%%%%%%%%%%%%%%%%%%%%%%%%%%%%%%%%
\section{Introduction}
%%%%%%%%%%%%%%%%%%%%%%%%%%%%%%%%%%%%%%%%%%%%%%%%%

The noncommutative geometry~\cite{connes}
remains among the strongest mathematical tools that can be used to
solve different problems in modern physics.
For instance, interesting results were reported for the quantum Hall effect~\cite{Prange} due either to the charge current~\cite{jellal}
or spin current~\cite{houca,dayi,ma,basu}. To remember, the noncommutative geometry is already exits and found its application in the fractional quantum Hall effect when the lowest Landau Level (LLL) is partially filled. It happened that in LLL, the potential energy is strong enough than kinetic energy and therefore the particles are glue in the fundamental level. As a consequence of this drastic reduction of the degrees of freedom, the two space coordinates  become noncommuting~\cite{jellal2} and satisfy the commutation relations analogue to those verifying by the position and the momentum in quantum mechanics.
 Also
various aspects of the
quantum  mechanics  have been
investigated in different ways in order to explore
the role of the noncommutative parameter in the physical observables
\cite{188,211, 222,244, 255}.

On the other hand, the noncommutative geometry has been employed to study different thermodynamics systems, one may see \cite{Huang, Hosseinzadeh, Shariati}. The main outcome is that  modification
of different thermodynamics quantities were obtained in terms of the noncommutative
parameter $\theta$.
 The Bose-Einstein condensation (BEC) was also taken
 part of the application of the noncommutative geometry. In fact,
 the thermodynamic properties of  BEC in the context
of the quantum field theory with non-commutative target space was studied in \cite{Brito}. 
Initially
BEC was theoretically predicted in 1924 \cite{ose} and  experimentally observed in 1995 \cite{Anderson}, which
 is a purely quantum phenomenon.
 Most
quantum effects occur either in the microscopic domain or at low temperatures. This condensation does not deviate from the rule
since it appears when one approaches the absolute zero $K$.

Motivated by different works mentioned above, we consider a  system of $N$ particles living on the noncommutative plane
and study its thermodynamics properties. In the first stage
we write the corresponding Hamiltonian using the star product definition
to end up with the solutions of the energy spectrum in terms of the noncommutative
parameter $\theta$. These will be used to explicitly determine the partition
function and therefore derive the related thermodynamics quantities such
the internal energy and heat capacity. In analogy with the magnetic field case, we
discuss the possibility of having an affective magnetization  with respect to $\theta$
and also getting the associated susceptibility. We also study BEC for the present
system and underline its main behavior. Finally interesting limiting cases
in terms of the involved parameters will be discussed
and some plots will be presented to give different illustrations
of our results.

The present paper is organized as follows. In section 2, we consider one particle
in 2-dimensions subjected to a harmonic potential and use the noncommuting coordinates
to end up with its noncommutative version.
This process allows us to end up with a Hamiltonian system similar to that of one particle living on the plane
in the presence of an external magnetic field. The corresponding energy spectrum will be given by using the algebraic approach through the annihilation
and creation operators. In section 3, we determine the partition function to end up
with  different thermodynamic quantities 
and study some limiting cases related to the temperature as well as $\theta$.
In section 4, we define an effective magnetization with respect to $\theta$
 and study
its susceptibility by considering some limits.
We analyze BEC
for the present system in section 5 and study its particular cases. We conclude our work in the final section.

%%%%%%%%%%%%%%%%%%%%%%%%%%%%%%%%%%%%%%%%%%%%%%%%%%%%%%%%%%%
\section{Solution of the energy spectrum}
%%%%%%%%%%%%%%%%%%%%%%%%%%%%%%%%%%%%%%%%%%%%%%%%%%%%%%%%%%%

We consider a particle of mass $m$ living on the plane $(x,y)$
and subjected to a confining potential. It is described by the Hamiltonian
%Our proposal can be elaborated by considering two decoupled harmonic oscillators
%having the same masses $m$ and frequencies $\om$. These are described by the Hamiltonian
\beq\lb{ccc}
H={1\over 2m}
\left(p_{x}^{2 }+p_{y}^{2}\right) +{m\om^2\over 2}\left(x^2+y^2\right)
\eeq
where   $\om$ is the frequency. To study the thermodynamics properties of
a system of $N$ particles described by (\ref{ccc}) on the noncommutative plane, we have to settle all ingredients needed to tackle our issues, which can be achieved
by
 adopting a method similar to
that used in~\cite{jellal}. Indeed, in addition the standard
canonical quantization between  the coordinate and momentum operators, we introduce
an algebra governed by the noncommutating coordinates
\beq \lb{aaa}
\left[x,y\right]=i\te
\eeq
where $\te$ is a real free parameter and has length square
of dimension.
Without loss of generality, hereafter we assume that $\te>0$ is fulfilled. From the above
consideration, we can now derive
 the noncommutative version of
the Hamiltonian (\ref{ccc}) as
\beq\lb{rrr}
H^{\sf nc}={1\ov2m_{\theta}} \left(p_x^{2}+p_y^{2}\right)
+{m\omega^{2}\over{2}}{\left(x^{2}+y^{2}\right)+{m\omega^{2}\over2\hbar}\te\left(yp_x-xp_y\right)}
\eeq
with the effective mass $m_\te= \frac{m}{1+\left(\frac{m\om\te}{2\hbar}\right)^2}$.

The obtained  Hamiltonian \eqref{rrr} can be diagonalized algebraically by introducing
the annihilation and creation operators
\begin{eqnarray}
&& a_x= {x\over\l_\te}+i{l_\te\over 2\hbar}p_x, \qquad a_x^{\da}= {x\over\l_\te}-i{l_\theta\over 2\hbar}p_x\\
&& a_y={y\over\l_\te}+i{l_\te\over 2\hbar}p_y, \qquad a_y^{\da}={y\over\l_\te}-i{l_\theta\over 2\hbar}p_y
\end{eqnarray}
satisfying the commutation relations
\beq
\left[a_x, a_x^{\da} \right]=\left[a_y, a_y^{\da} \right]= \mathbb{I}
\eeq
with the noncommutative length $
l_\te={\sqrt[4]{ \left(\frac{2{\hbar}}{m\om}\right)^2+{\te}^{2}}}.$ Another set of
operators can be defined, such as
\begin{eqnarray}
&& a_d={ 1\over\sqrt{2}}({a_x-ia_y}), \qquad a_d^\da={ 1\over\sqrt{2}}({a_x^\da+ia_y^\da})\\
&& a_g={ 1\over\sqrt{2}}(a_x+ia_y), \qquad a_g^\da={ 1\over\sqrt{2}}(a_x^\da-ia_y^\da)
\end{eqnarray}
verifying
\beq
\left[a_d, a_d^{\da} \right]=\left[a_g, a_g^{\da} \right]= \mathbb{I}
\eeq
and all other commutators are vanishing. Now
combining all and using
 the operator numbers $N_d=a_d^{\da}a_d$ and
$N_g=a_g^{\da}a_g$, to write the Hamiltonian \eqref{rrr} as
\beq\lb{ncham}
H^{\sf nc}={m\omega^{2}\over{2}}(l_\te^{2}+\te)N_d+{m\omega^{2}\over{2}}(l_\te^{2}-\te)N_g+{m\omega^{2}\over{2}}{l_\te^{2}}
\eeq
which has the following solution of the energy spectrum
\begin{eqnarray}
&&
\lb{rerr}
E_{n_d,n_g}= {m\omega^{2}l_\te^2\over{2}}(n_d+n_g +1) + {m\omega^{2}\te\over{2}}(n_d - n_g)
\\
&&
|n_d,n_g\rangle= {(a^{\dagger}_d)^{n_d}(a^{\dagger}_g)^{n_g}\over
  \sq{(n_d!)(n_g!)}} |0,0\rangle, \qquad n_d,n_g\in \mathbb{N}.
\end{eqnarray}

In the next we will show how the above results can be used
to  investigate the main thermodynamics features
of the present system.

%%%%%%%%%%%%%%%%%%%%%%%%%%%%%%%%%%%%%%%%%%%%%%%%%%%%%%%%%%%%%%%%%%%%%%%%%%%%%%%%%
\section{Thermodynamics quantities}
%%%%%%%%%%%%%%%%%%%%%%%%%%%%%%%%%%%%%%%%%%%%%%%%%%%%%%%%%%%%%%%%%%%%%%%%%%%%%%%%%%

As usual to determine different thermodynamics quantities, one has to start from
the corresponding partition function
\beq\lb{dfpa}
\mathbb{Z}= \Tr e^{-\beta H}
\eeq
where $\beta=\frac{1}{k_BT}$, $k_B$ the Boltzmann constant, $T$ the temperature and
$H$ is the Hamiltonian for a given system. In terms of the above solution
of the energy spectrum, \eqref{dfpa} takes the form
\beq
\mathbb{Z}_\theta=\sum_{n_d=0}^\infty\sum_{n_g=0}^\infty e^{-\beta E_{n_d,n_g}}.
\eeq
To proceed further, let us  rearrange the eigenvalues (\ref{rerr})
as
\beq
 E_{n_d,n_g}=\left(\phi_\theta+\varphi_\theta\right)n_d+\left(\phi_\theta-\varphi_\theta\right)n_g+\phi_\theta
\eeq
by involving  two parameters $\theta$-dependent $\phi_\theta = {m\omega^2l_\theta^2\ov2}$ and $\varphi_\theta = {m\omega^2\theta\ov2}$.
After  straightforward calculation, we end up with the partition function for one
particle
\beq
\mathbb{Z}_\theta={1\ov2\cosh\left(\beta \phi_\theta\right)-2\cosh\left(\beta \varphi_\theta\right)}.
\eeq
It is clearly seen that
for a system of $N$ non-interacting particles, the total partition function is simply
given by the product
\beq\lb{Z_T}
\mathbb{Z}_T=\left({1\ov2\cosh\left(\beta \phi_\theta\right)-2\cosh\left(\beta \varphi_\theta\right)}\right)^{N}
\eeq
which is actually depending on two parameters of our theory, temperature and
noncommutative parameter. These will be used to study different limiting cases  and therefore characterize the present system behavior.

Having obtained all ingredients needed, now
%We have obtained the principal results and now
we can determine different thermodynamics quantities related to the present system. Indeed, as far as the internal energy is concerned we start from the usual definition
\beq
U=-{\partial \ln \mathbb{Z}_T\ov\partial\beta}
\eeq
to end up with the form
%and using (\ref{Z_T}) to obtain
\beq
U_\theta=N{\phi_\theta \sinh\left(\beta \phi_\theta\right)-\varphi_\theta \sinh\left(\beta \varphi_\theta\right)\ov \cosh\left(\beta \phi_\theta\right)-\cosh\left(\beta \varphi_\theta\right)}.
\eeq
We notice that
there are two limiting cases that can be considered  with respect to
the noncommutative parameter
$\theta$. Indeed, firstly by switching off $\theta$, we recover the standard form
\beq
U_{\theta=0}= 2Nk_BT\hbar\omega{\sinh\left(\beta\hbar\omega\right)\ov \cosh\left(\beta\hbar\omega\right)-1}
\eeq
and 
secondly, by requiring the limit $\theta \longrightarrow 0$, we end up with a linear behavior
in terms of temperature
\beq\lb{th00}
U_{\theta\longrightarrow 0}\simeq2Nk_BT.%\hbar\omega\left({\sinh\left(\beta\hbar\omega\right)\ov \cosh\left(\beta\hbar\omega\right)-1}\right).
\eeq
Note that, \eqref{th00}
is independent of the noncommutative parameter, a  result that will be confirmed
in the next analysis.

To characterize  thermally the present system let us consider the heat capacity. Then from the above result and using the relation
\beq
C_{\theta}={\partial U_\theta\ov \partial T}
\eeq
one gets the following heat capacity
\beq
C_{\theta}=-Nk_B\beta^2 \frac{\phi_\theta^2 \cosh (\phi_\theta \beta )-\varphi_\theta^2 \cosh (\beta  \varphi_\theta)}{\cosh (\phi_\theta \beta )-\cosh (\beta  \varphi_\theta)}-\frac{(\phi_\theta \sinh (\phi_\theta \beta )-\varphi_\theta \sinh (\beta  \varphi_\theta))^2}{(\cosh (\phi_\theta \beta )-\cosh (\beta  \varphi_\theta))^2}
\eeq
which can be studied according to different limits taken be
%After this stage we can inspect some special cases related to
by the parameters $\theta$ and $\beta$. Indeed, for $\theta=0$, we recover
the usual result
\beq
C=Nk_B{(\beta\hbar\omega)^2\ov \cosh\left(\beta\hbar\omega\right)-1}
\eeq
and at high temperature limit, it reduces to the quantity
\beq
C_{\theta}=2Nk_B\left(1- \frac{1}{2} \beta ^2 ({\varphi_\theta^2+\phi_\theta^2})\right)
\eeq
which is showing an extra term removed from the standard result $2Nk_B$ that can be interpreted as
a quantum correction to the heat capacity. This result might be interesting in dealing with
the vibration of atoms in solid state physics or other systems in order
to give a laboratory test of the noncommutative parameter.
However, at low temperature  we show that $C_{\theta}$ vanishes that is in agreement with the standard result.

\begin{figure}[H]
\centering{
\includegraphics[width=12cm,height=8cm]{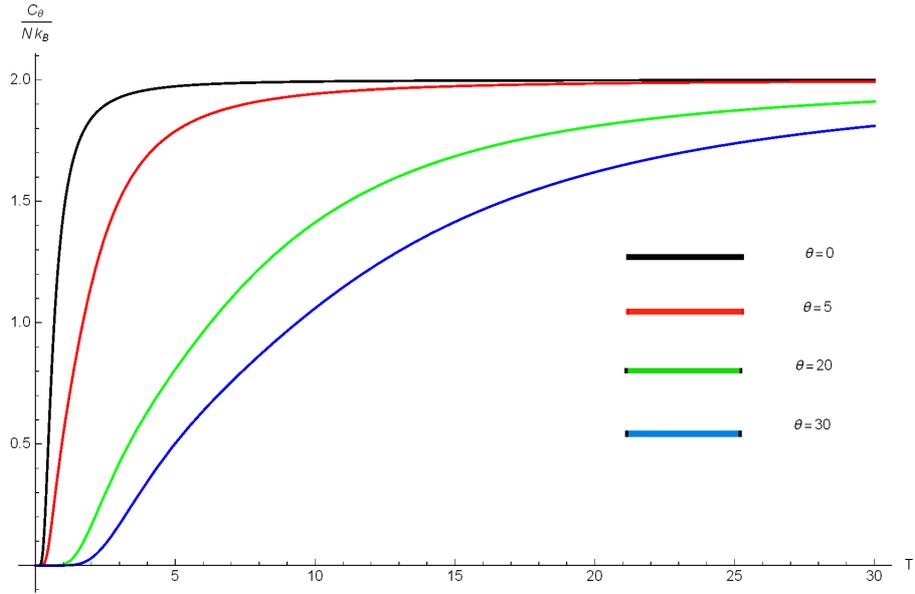} %\hfill
\caption{\sf (Color online) The heat capacity versus the temperature $T$ for four values of the
noncommutative parameter $\theta=0, 5, 20, 30$.}
\label{CV0h}}
\end{figure}

 Figure \ref{CV0h} presents the heat capacity $C_{\theta}$ as function
 of the temperature
 for four values of the noncommutative parameter $\theta=0, 5, 20, 30$. We observe that $C_{\theta}$  increases quickly toward a constant value in terms of $T$ when $\theta$ is small. However
when  $\theta$ becomes large and even  $T$ increases, the heat capacity remains null,
which causes a change of its origin. This behavior
%This shows a depalcement of the $C_{\theta}$  orgin and also
tells us that
there a threshold value $T_s (\theta)$ of the temperature, which is
$\theta$-dependent. Thus, we conclude that
if  $T<T_s (\theta)$ the heat capacity remains null, while it  increases
to reach constant values according to the fixed values taken by $\theta$ when  the condition
is fulfilled $T>T_s(\theta)$.
On the other hand, at high temperature $C_\theta$ vanishes when $T$ takes the form
\beq
T(\theta)={m\omega^2\ov2k_B}\sqrt{\theta^2+2\left({\hbar\ov m\omega}\right)^2}
\eeq
which can be used to give a measurement of the noncommutative parameter through
the temperature variation and therefore argue the validity of introducing the noncommutating
coordinates in the present system.

\begin{figure}[H]
\centering{
\includegraphics[width=12cm,height=8cm]{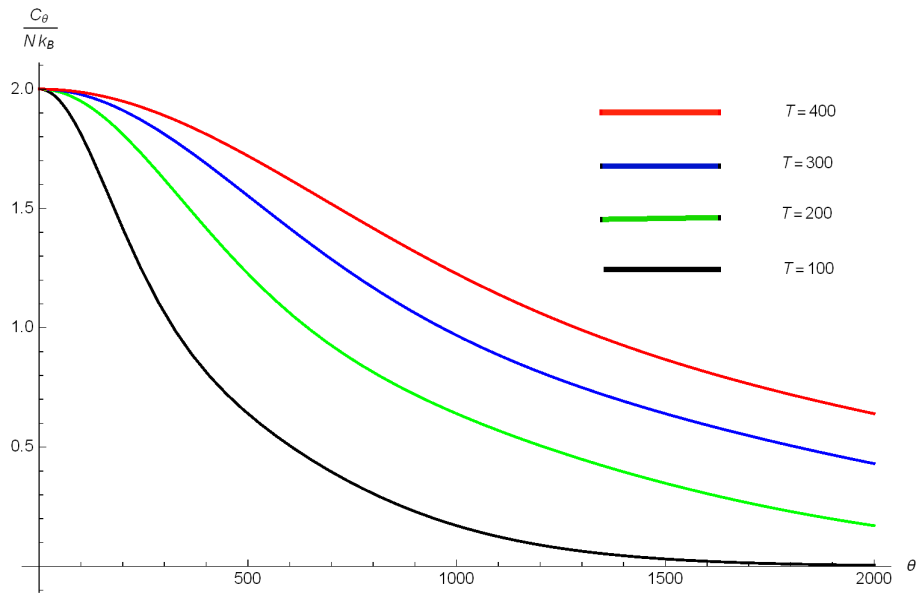} %\hfill
\caption{\sf (Color online) The heat capacity versus the
noncommutative parameter $\theta$ for four values of the temperature
$T=100 K, 200 K, 300 K, 400 K$.}
\label{CV0h2}}
\end{figure}

To accomplish such numerical analysis of the heat capacity $C_{\theta}$,
 in Figure \ref{CV0h2} we plot  $C_{\theta}$ in terms of   the noncommutative parameter
 $\theta$ for different values of temperature $T=100 K, 200 K, 300 K, 400 K$. It is clearly seen that $C_{\theta}$
 decreases rapidly for some values of the temperature but such behavior changes
 once $T$ increases giving rise different results. Thus,
 we conclude that the heat capacity can be controlled by changing  $\theta$
 together with the temperature.

%%%%%%%%%%%%%%%%%%%%%%%%%%%%%%%%%%%%%%%%%%%%%%%%%
\section{Effective magnetization}
%%%%%%%%%%%%%%%%%%%%%%%%%%%%%%%%%%%%%%%%%%%%%%%%%%

Recall that
the present study does not include an external magnetic field but  we can still talk about
magnetization since $\theta$ is a free parameter.  field.
Then in analogy, we can define an effective magnetization in the same
as for the case of a magnetic field and write
\beq
M_\theta={1\ov\beta}{\partial \ln {\mathbb Z}_T\ov\partial\theta}.
\eeq
After calculation, we end up with
\beq
M_\theta={Nm\omega^2\ov2}{\sinh(\beta\varphi_\theta)-
{\theta\ov l_\theta^2}\sinh(\beta\phi_\theta)\ov \cosh(\beta\phi_\theta)-\cosh(\beta\varphi_\theta) }
\eeq
which is depending on $\theta$ as well as the temperature and therefore one can study some limiting cases
to underline its behavior. Indeed, at low temperature $(\beta\longrightarrow\infty)$, $M_\theta$ becomes
\beq\lb{Mxx}
M_\theta=-\frac{Nm\omega^2} {2 {\sqrt{ 1+\left(\frac{2{\hbar}}{m\om\theta}\right)^2}}}
\eeq
and at high temperature $(\beta\longrightarrow 0)$, we obtain a linear dependence in terms of $\theta$
\beq\lb{Mx}
M_\theta=-\frac{\beta N m^2\omega^4}{24}\theta.
\eeq

\begin{figure}[H]
\centerline{\includegraphics[width=12cm,height=8cm]{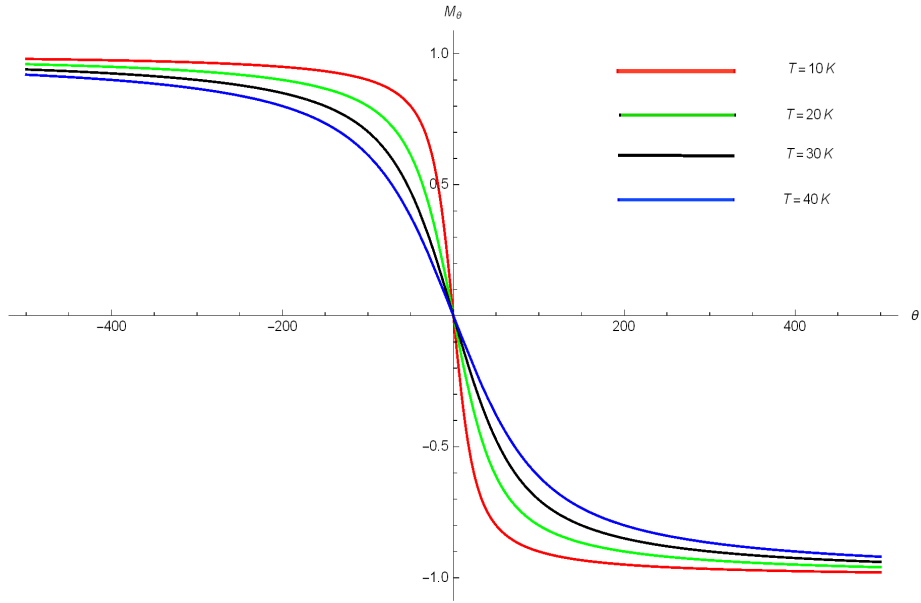}}
\caption{(Color online) The effective magnetization $M_\theta$ in terms of
the noncommutative parameter $\theta$ for different values of the temperature $T=10 K, 20 K, 30 K, 40 K$.}
\label{MoT}
\end{figure}

In Figure \ref{MoT}, we plot the effective magnetization versus the noncommutative parameter $\theta$
for different values of the temperature $T$.
A very important point is that when the parameter $\theta$ is weak for high or low temperature, $M_\theta$
varies in a linear way. % according to (\ref{Mx}).
However for the case when $\theta$ is strong in low temperature $M_\theta$ becomes constant. This tells us that
one may use such magnetization to measure the noncommutative
parameter.

At this level, we can also introduce the effective
%Now we can introduce the
susceptibility by adopting that corresponding to the magnetic field and thus have
\begin{eqnarray}
  \chi_\theta={\partial M_\theta\ov \partial\theta} &=& \frac{m N \omega ^2}{4l_{\theta }^6}\
  \frac{{2\theta ^2 \sinh \left(\beta  \phi _{\theta }\right)} %{}
  -{2l_{\theta }^4\sinh \left(\beta  \phi _{\theta }\right)} %{l_{\theta }^2}
  -{l_{\theta }^2\beta  \theta ^2 m \omega ^2 \cosh \left(\beta  \phi _{\theta }\right)}
  %{2 l_{\theta }^4}+\frac{1}{2}
  +l_{\theta }^6 \beta  m \omega ^2 \cosh \left(\beta  \varphi _{\theta }\right)}
  {\cosh \left(\beta  \phi _{\theta }\right)-\cosh \left(\beta  \varphi _{\theta }\right)} \nonumber \\
   && - \frac{\left(l_{\theta }^2\sinh \left(\beta  \varphi _{\theta }\right)-
   {\theta  \sinh \left(\beta  \phi _{\theta }\right)}\right)
   \left({\beta  \theta  m \omega ^2 \sinh \left(\beta  \phi _{\theta }\right)}
   - l_{\theta }^2 \beta  m \omega ^2 \sinh \left(\beta  \varphi _{\theta }\right)\right)}
   {2l_{\theta }^4\left(\cosh \left(\beta  \phi _{\theta }\right)-\cosh \left(\beta  \varphi _{\theta }\right)\right){}^2}.
\end{eqnarray}
At high temperature $\beta\longrightarrow 0$,
 it can be approximated  as
\beq\lb{para}
\chi_\theta=-\frac{\beta N m^2 \omega ^4}{12}
\eeq
which is similar to the well-known Curie law ${C\ov T}$ where
the Curie constant can be fixed as 
$C=-{N m^2\omega^4\ov 12k_B}$. It is clearly seen that
%From this result,
from \eqref{para},
%This tells us that
%we conclude that
the present system behaves like a diamagnetism and also 
like a superconductor  if we require
\beq
T_{SC}=\frac{12k_B }{N m^2\omega^4}.
\eeq
These show that our results are general and can be tuned on to give
different interpretations of the present system. On the other hand,
%recover different standard results
%and therefore
one can use them to give a laboratory test of the
noncommutative parameter.

%%%%%%%%%%%%%%%%%%%%%%%%%%%%%%%%%%%%%%%%%%%%%%%%%
\section{Bose-Einstein condensation }
%%%%%%%%%%%%%%%%%%%%%%%%%%%%%%%%%%%%%%%%%%%%%%%%%%

Let us investigate the relation between the corresponding Bose-Einstein condensation
%Her using all ingredients,  we can now  study the Bose-Einstein condensation
and the noncommutative parameter $\theta$.
For this we start by introducing the quantities
\beq
\omega_1 = {m\omega^2l_\theta^2\ov2}+{m\omega^2\theta\ov2} ,\qquad \omega_2 =
{m\omega^2l_\theta^2\ov2}-{m\omega^2\theta\ov2},\qquad \varepsilon_0 = {m\omega^2l_\theta^2\ov2}
\eeq
 to rewrite the single-particle energy  as
\beq
E_{n_gn_d}=\varepsilon+\varepsilon_0=\omega_1n_d+\omega_2 n_g+\varepsilon_0.
\eeq
Unlike the case of fermions, any number of bosons may be placed in a particular
micro-state and when the temperature of the system is null, all bosons must be in the ground-state energy $\varepsilon_0$. Such
%The latter being equal to only the
number of bosons, that are not in the ground-state  $\varepsilon_0$, 
is given by
%$N_{\varepsilon>0}$  be this number, such as we have
\beq
N_{\varepsilon}=\sum_{n_g,n_d=0}^{+\infty}{1\ov \exp\left({E_{n_gn_d}-\mu\ov k_BT}\right)-1}
\eeq
where $\mu$ is the chemical potential. Then immediately, we derive
the number of bosons in the ground-state $\varepsilon_0$ %is, for its part, equal to
\beq\lb{mm}
N_{\varepsilon=0}=N-N_{\varepsilon>0}.
\eeq
Note that, when $T$ is high we have $N_{\varepsilon=0}\ll N_{\varepsilon>0}$, 
but at $T = 0$
this is no longer the case because all  particles are in the corresponding micro-state
$\varepsilon = 0$
and therefore we can write
\beq\lb{aaa1}
\langle N_{\varepsilon=0}\rangle=N={1\ov \exp\left({\varepsilon_0-\mu\ov k_BT}\right)-1}.
\eeq

Now we adopt a procedure similar to  that applied in \cite{Grossmann,Gros} in order to
determine the proper density of states $\rho(\varepsilon)$. This latter
%Now we choose different strategies beginning by replace the summation by integral and searching the proper density of states, similar to the procedure applied in \cite{Grossmann}. The desired density  $\rho(\varepsilon)$
can be obtained from the number $\nu(\varepsilon)$ of states for which $\varepsilon=\omega_1n_d+\omega_2 n_g$
is less than or equal to a given energy. To describe the Bose-Einstein condensation we restrict our self to the case where $\theta$ is small
and therefore
we can make an expansion
to write a
new frequency $\Omega=\omega_1\omega_2$ and the ground-state energy as
\beq
\Omega^2=\omega^2\left(1-{\theta^2\ov\alpha^2}\right),\qquad \varepsilon_0 = \hbar\omega\left(1+{\theta^2\ov2\alpha^2}\right)
\eeq
which allow to end up with
the desired density $\rho(\varepsilon)$
\beq\lb{rho}
\rho(\varepsilon)=\frac{1}{\hbar\omega  \sqrt{1-\frac{\theta ^2}{\alpha ^2}}}+
\frac{\varepsilon }{\left(\hbar\omega\right)^2 \left(1-\frac{\theta ^2}{\alpha ^2}\right)}
\eeq
where we have defined the parameter $\al=\frac{2\hbar}{m\om}$
that has to fulfill the condition $\al>\theta$.
To go further, we
replace $\omega_1n_d+\omega_2 n_g$ by a continuous variable $\varepsilon$
and consider  (\ref{rho}) to convert the summation over the quantum numbers
$n_d, n_g$ on integration over $\varepsilon$. Doing this process to obtain
\beq\lb{Ni}
N_{\varepsilon>0}=\int_0^{+\infty}{\rho(\varepsilon)\ov
\exp\left({\varepsilon+\varepsilon_0-\mu\ov k_BT}\right)-1}d\varepsilon
\eeq
which can be written in terms of
 the fugacity $z_\theta=\exp\left({\mu-\varepsilon_0\ov k_BT}\right)$ %to write (\ref{Ni})
 as
\beq\lb{Nc}
N_{\varepsilon>0}=\frac{1}{\left(\hbar\omega\right) ^2 \left(1-\frac{\theta ^2}{\alpha ^2}\right)}
\int_0^{+\infty}{\varepsilon\ov z_\theta^{-1}\exp\left({\varepsilon\ov k_BT}\right)-1}d\varepsilon+\frac{1}
{\hbar\omega \sqrt{\left(1-\frac{\theta ^2}{\alpha ^2}\right)}}\int_0^{+\infty}
{1\ov z_\theta^{-1}\exp\left({\varepsilon\ov k_BT}\right)-1}d\varepsilon
\eeq
or equivalently
\beq\lb{Ne}
N_{\varepsilon>0}=\frac{\left(k_BT\right)^2}{\left(\hbar\omega\right) ^2 \left(1-\frac{\theta ^2}
{\alpha ^2}\right)}\int_0^{+\infty}{x\ov z_\theta^{-1}e^x-1}dx+\frac{k_BT}{\hbar\omega \sqrt{\left(1-\frac{\theta ^2}
{\alpha ^2}\right)}}\int_0^{+\infty}{1\ov z_\theta^{-1}e^x-1}dx
\eeq
after making the change of variable $x={\varepsilon\ov k_BT}$.
%Introducing the Bose-Einstein function \cite{Pathria}
%\beq\lb{ggg}
%g_n(z)={1\ov\Gamma(n)}\int_0^{+\infty}{x^{n-1}\ov z^{-1}e^{x}-1}dx
%\eeq
We show that both of
%It is clearly seen that both of
integrals converge only when the fugacity is in the interval $0<z_\theta<1$ %\in]0,1[$,
%i.e. $-\infty<\mu<\varepsilon_0$,
and therefore we derive a strong condition
to  obtain the Bose-Einstein condensation that is the noncommutative
parameter has to satisfy the following restriction
\beq
 \alpha  \sqrt{2\left|\frac{\mu }{\hbar \omega }-1\right|}<\theta<\al.
\eeq
With this, we can now integrate
(\ref{Ne}) to get
%\beq
%g_2(1)={1\ov\Gamma(2)}\int_0^{+\infty}{x\ov e^{x}-1}dx={\pi^2\ov6}
%\eeq
%and finally, we get  the expression %of $N$ when
\beq\lb{kf}
N_{\varepsilon>0}=\frac{\left(k_BT\right)^2}{\left(\hbar\omega\right) ^2 \left(1-\frac{\theta ^2}
{\alpha ^2}\right)}\text{Li}_2\left(z_{\theta }\right)-\frac{k_BT}{\hbar\omega
\sqrt{\left(1-\frac{\theta ^2}{\alpha ^2}\right)}}\ln \left(1-z_{\theta }\right)
\eeq
where $\text{Li}$ is  the polylogarithm function. Recall that the Bose temperature $T_C$ is that
for which $N_{\varepsilon>0}=N$ and then we have
%A condensation temperature $T_C$ can now be introduced, in fact $T_C$ as that the temperature where  the ground state population $N_0$ should become zero then
%the equation  (\ref{kkk}) give
\beq\lb{kkk}
T_C={\hbar\omega\ov k_B} \sqrt{\left(1-{\theta^2\ov\alpha^2}\right)}\left(\frac{N}{\text{Li}_2\left(z_{\theta }\right)}\right)^{1/2} \left(k_B  T_C \ln \left(1-z_{\theta }\right)\ov \hbar\omega \sqrt{\left(1-\frac{\theta ^2}{\alpha ^2}\right)}N +1\right)^{1/2}.
\eeq
Now if the number of particles is very large, then the Bose-Einstein condensation occurs at the temperature
\beq\lb{000}
T_0={\hbar\omega\ov k_B} \sqrt{\left(1-{\theta^2\ov\alpha^2}\right)}\left(\frac{N}{\text{Li}_2
\left(z_{\theta }\right)}\right)^{1/2}.
\eeq
Using (\ref{mm}) and (\ref{000}) to obtain the ratio
\beq
\frac{N_0}{N}=1-\left(\frac{T}{T_0}\right)^2+\frac{\ln \left(1-z_{\theta }\right)}{\sqrt{N}
\text{Li}_2\left(z_{\theta }\right)}\left(\frac{T}{T_0}\right).
\eeq

\begin{figure}[h]
\centering{
\includegraphics[width=12cm,height=8cm]{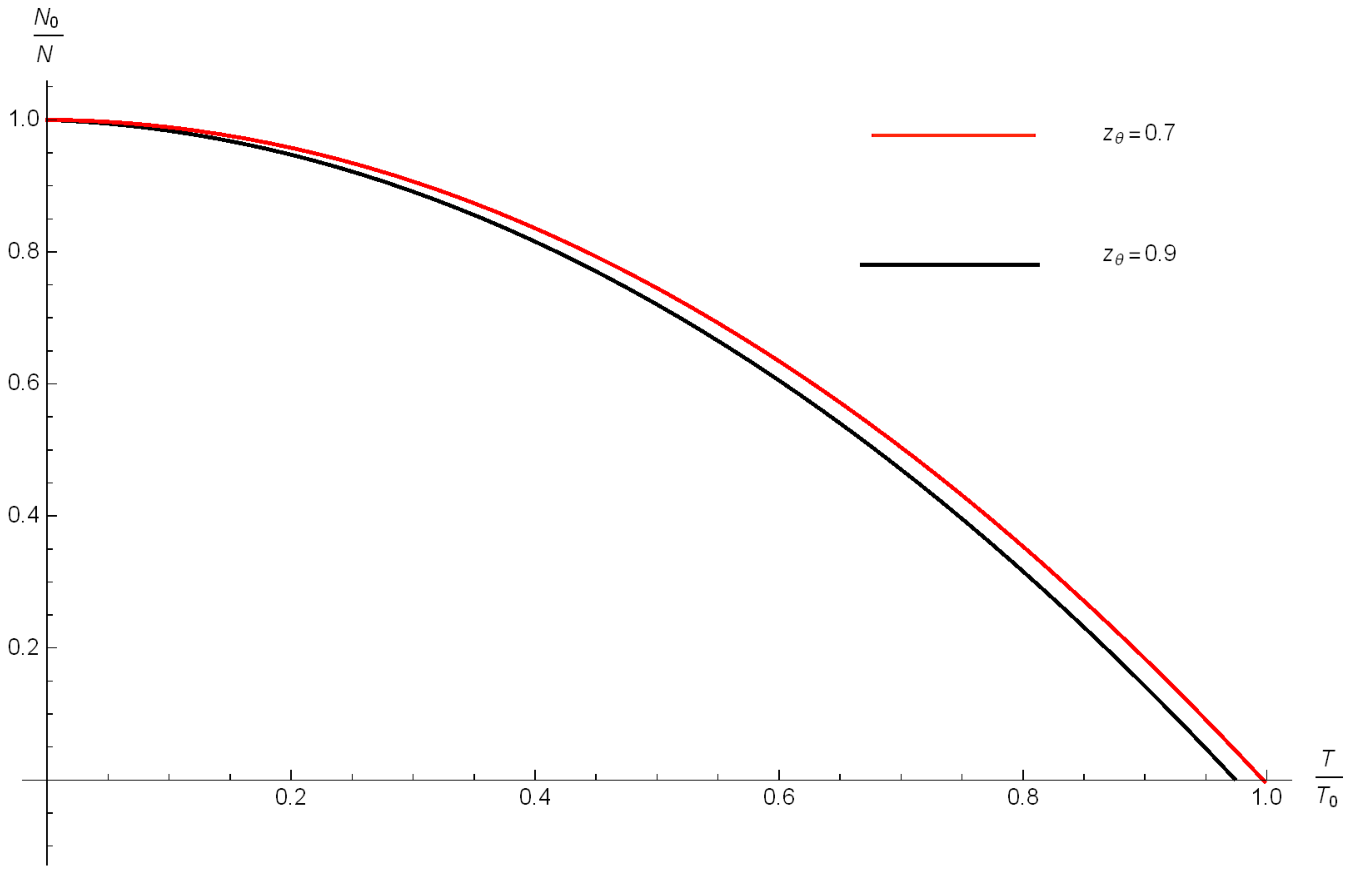}\\
\ \ \includegraphics[width=12cm,height=8cm]{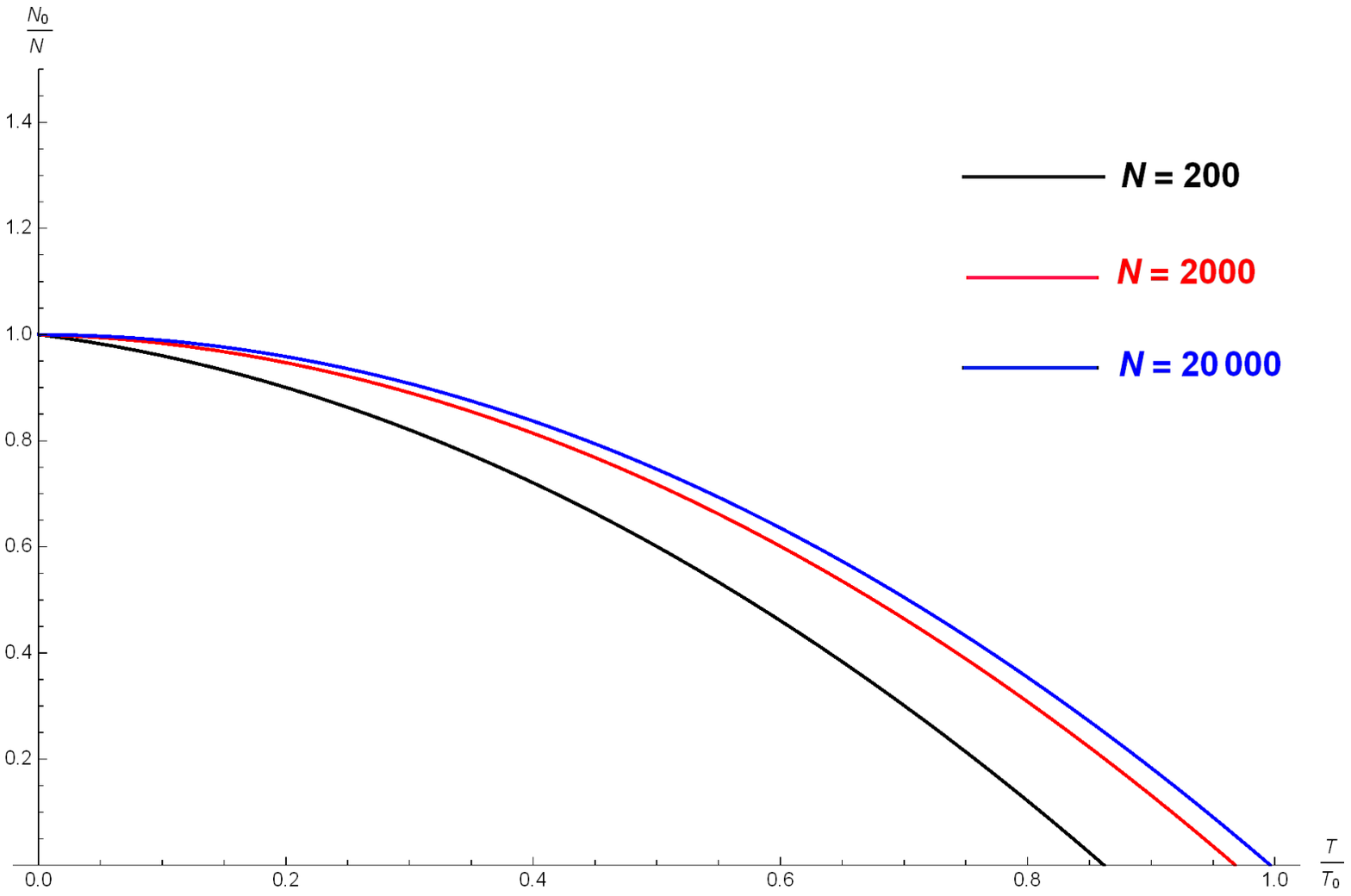}}
\caption{(Color online) The number ratio ${N_0\ov N}$ versus the temperature ratio ${T\ov T_0}$ for some values of the fugacity.}
\label{bose}
\end{figure}
In Figure \ref{bose} we plot the number ratio ${N_0\ov N}$ versus the temperature ratio ${T\ov T_0}$ for some values of the fugacity.
%It decreases as long as the temperature increases and becomes
%null once both temperatures coincide.
The shape of the Bose-Einstein condensation becomes spherical when increasing the fugacity or decreasing
the number of particles, which are two   parameters  governing the ellipticity of the shape of the Bose-Einstein condensation.
These are interesting remarks because in the physics we know that the elliptical shape is a consequence
of the precise geometry of the trap in which the superfluid is maintained. Thus one can modify such shape by changing
the magnetic field that creates this trap. Then in our study we can do the same job by fixing the noncommutative parameter as a magnetic
field and then  modify the shape of the Bose-Einstein condensation.

Let us  look for the relation between the noncommutative parameter and the condensation
temperature $T_C$ to underline the behavior of the present system. This can be done
by considering (\ref{kf})  to show that such relation
is given by
%to write  the explicit relation between the Bose temperature and the noncommutative parameter
\beq\lb{ww}
T_C=\frac{\sqrt{4 N \text{Li}_2\left(z_{\theta }\right)+\log ^2\left(1-z_{\theta }\right)}+
\log \left(1-z_{\theta }\right)}{2  \text{Li}_2\left(z_{\theta }\right)}.
\eeq

\begin{figure}[H]
\centerline{\includegraphics[width=12cm,height=8cm]{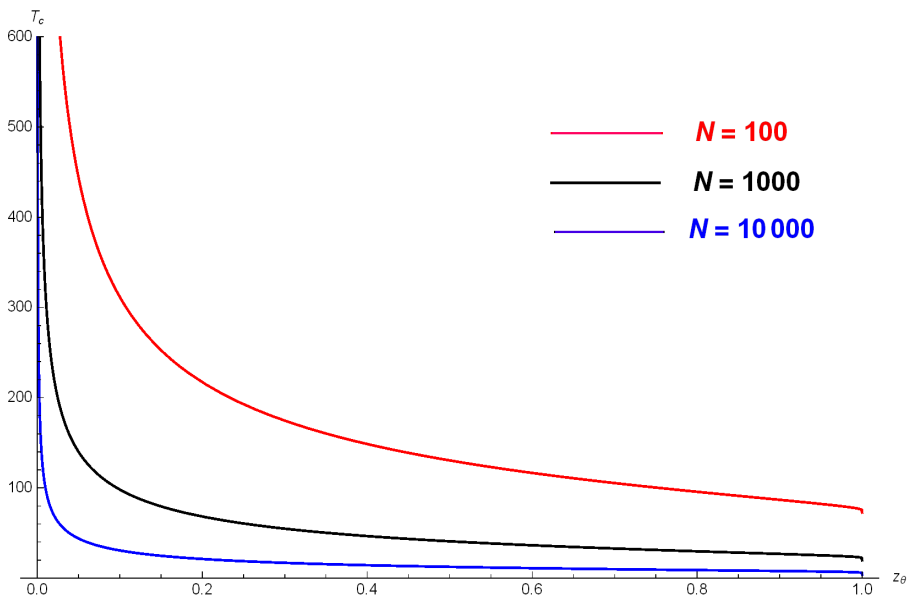}}
\caption{(Color online) The condensation temperature $T_C$ versus the fugacity $z_\theta$ for some 
particular values of $N$.}
\label{TC}
\end{figure}

Figure \ref{TC}  shows that the temperature of condensation is strongly depending  on the
fugacity, which is a function of the noncommutative parameter $\theta$, together with
the number of particles.
We observe that when the fugacity is close to zero $T_c$
decreases rapidly but when the fugacity increases to move away from zero $T_c$
remains almost constant.
On the other hand, $T_c$ tends to zero when $N$ becomes of the order of $10^4$ but if $N$ is of the order of $10^3$, $T_c$ tends to a non-null value.

%%%%%%%%%%%%%%%%%%%%%%%%%%%%%%%%%%%%%%%
\section {Conclusion}
%%%%%%%%%%%%%%%%%%%%%%%%%%%%%%%%%%%%%

We have studied the thermodynamic properties and analyzed  the Bose-Einstein condensation 
for a system of $N$ particle living on the noncommutative plane.
%The condition on $\theta$ to obtain it, for that we have employed two decoupled. 
After building the noncommutative Hamiltonian via star product definition and
 getting the solution of the energy spectrum, 
we have determined the partition function
 $Z_\theta$ in terms of the noncommutative
parameter $\te$. This was used to  derive  the corresponding  internal energy
 and therefore the heat capacity.

Subsequently, we have defined an effective magnetization %$M_\theta$ 
in similar way to that corresponding
to the magnetic field. It was noticed that
 when the parameter $\theta$ is very low for high or low temperature regimes,
the effective magnetization %$M_\theta$
varies in a linear way. On the other hand, by evaluating the associated 
susceptibility, we have obtained a negative expression at high temperature, which showing   similarity with the Curie law for a magnetic system.
Finally,
we have shown that to get the Bose-Einstein condensation in the present system, one has to 
fix the noncommutative parameter
 $\te$ in a well-defined interval. This was used to establish 
 an interesting relation between  the temperature of condensation and  $\theta$.

 %, if not the system will not have a Bose-Einstein condensation.

%%%%%%%%%%%%%%%%%%%%%%%%%%%%%%%
\section*{Acknowledgment}
%%%%%%%%%%%%%%%%%%%%%%%%%%%%%%

The generous support provided by the Saudi Center for Theoretical
Physics (SCTP) is highly appreciated by all authors.

%%%%%%%%%%%%%%%%%%%%%%%%%%%%%%%%%%%%%%%%%%%%%%%%%%%%%%%%%%%%%%%%%%%%%%%%%%%%%%%%%%%%%%%%%%%%%%%%%%%%%%%%%%%%%%%%%%%%%%%%%%%%%%%%%%%%%%%%%%%%%%%%%%%%%%%%%%%%%%%%%%%%%%%%%%%%%%%%%%%%%%%%%%%%%%%%%%%%%%%%%%%%%%%%%%%%%%%%%%%%%%%%%%%%%%%%%%%%%%%%%%%%%%%%%%
%%%%%%%%%%%%%%%%%%%%%%%%%%%%%%%%%%%%%%%%%%%%%%%%%%%%%%%%%%%%%%%%%%%%%%%%%%%%%%%%%%%%%%%%%%%%%%%%%%%%%%%%%%%%%%%%%%%%%%%%%%%%%%%%%%%%%%%%%%%%%%%%%%%%%%%%%%%%%%%%%%%%%%%%%%%%%%%%%%%%%%%%%%%%%%%%%%%%%%%%%%%%%%%%%%%%%%%%%%%%%%%%%%%%%%%%%%%%%%%%%%%%%%%%%%


\begin{thebibliography}{99}



\bibitem{connes} A. Connes, Noncommutative Geometry, (Academic Press 1994).


\bibitem{Prange} R. E. Prange and S. M. Girvin (editors), The Quantum Hall Effect, (SpringerVerlag 1990).

\bibitem{jellal} O. F. Dayi and A. Jellal, J. Math. Phys. 43, 4592 (2002). % 4592, hep-th/0111267

\bibitem{houca} A. Jellal and R. Hou\c{c}a, Int. J. Geom. Meth. Mod. Phys. 6, 343 (2009).

\bibitem{dayi} O. F. Dayi and M. Elbistan, Phys. Lett. A 373, 1314 (2009).

\bibitem{ma} K. Ma and S. Dulat, Phys. Rev. A 84, 012104 (2011).

\bibitem{basu} B. Basu, D. Chowdhury and S. Ghosh, Phys. Lett. A 377, 1661 (2013).
%Inertial spin Hall effect in noncommutative space, arXiv:1212.4625.

\bibitem{jellal2} A. Jellal and M. Bellati,  Int. J. Geom. Meth. Mod. Phys. 7, 143 (2010).



\bibitem{188}
J. Gamboa, M. Loewe, F. Mendez and J. C. Rojas,
%“Noncommutative quantum mechanics,”
Phys. Rev. D 64, 067901 (2001).

%\bibitem{199}
%J. M. Romero, J. A. Santiago and J. D. Vergara,
%“Newton’s second law in a non-commutative space,”
%Phys. Lett. A 310, 9 (2003).

%\bibitem{200}
%M. Daszkiewicz and C. J. Walczyk,
%“Newton equation for canonical, Lie-algebraic, and quadratic deformation of classical space,”
%Phys. Rev. D 77, 105008 (2008).

\bibitem{211}
J. Jing, F. H. Liu and J. F. Chen,
%“Classical and quantum mechanics in the generalized non-commutative plane,”
Europhys. Lett. 84, 61001 (2008).

\bibitem{222} B. Muthukumar and P. Mitra,
%“Noncommutative oscillators and the commutative limit,”
Phys.  Rev.  D 66,  027701 (2002).

%\bibitem{233}
%A. Kijanka and P. Kosinski,
%“Noncommutative isotropic harmonic oscillator,”
%Phys.  Rev. D 70, 127702 (2004).

\bibitem{244}
J. Jing, S. H. Zhao, J. F. Chen and Z. W. Long,
%“On the spectra of noncommutative 2D harmonic oscillator,” The European
Euro. Phys. J. C 54, 685 (2008).

\bibitem{255}
 A. Das, H. Falomir, M. Nieto, J. Gamboa and F. Mendez, ´
%“Aharonov-Bohm effect in a class of noncommutative theories,”
Phys. Rev. D 84, 045002 (2011).

\bibitem{Huang} Wung-Hong Huang and Kuo-Wei Huang,
	Phys. Lett. B 670, 416 (2009).

\bibitem{Hosseinzadeh} V. Hosseinzadeh, M. A. Gorji, K. Nozari and B. Vakili,
Phys. Rev. D 92,  025008 (2015).

\bibitem{Shariati} Ahmad Shariati, Mohammad Khorrami and Amir H Fatollahi,
J. Phys. A: Math. Theor. 43,  285001 (2010).



\bibitem{Brito} Francisco A. Brito and Elisama E.M. Lima,
Int. J. Mod. Phys. A 31, 1650057 (2016).

\bibitem{ose} S. Bose, Z. Phys. 26, 178 (1924).

\bibitem{Anderson}
M. H. Anderson, J. R. Ensher, M. R. Mattews,  C. E. Wieman and
	E. A. Corneli, Science 269, 198 (1995).
	


%\bibitem{Bagnato} V. Bagnato, D. E. Pritchard and D. Kleppner, Phys. Rev. A 35, 4354   %(1987).


%\bibitem{Chudnovsky} E. M. Chudnovsky, Phys. Rev. Lett. 99, 206601 (2007).




\bibitem{Grossmann} S. Grossmann and M. Holthaus, Z. Phys. B 97, 319  (1995).

\bibitem{Gros} S. Grossmann and M. Holthaus, Z. Natureforsch. 50 a, 921-930 (1995).

%\bibitem{Pathria} R. K. Pathria, statistical Mechanics Pergamon, Oxford (1985).


%\bibitem{gazeau} J. P. Gazeau, P. Y. Hsiao and A. Jellal, Phys. Rev. B 65,   094427 (2002).


\end{thebibliography}
\end{document}